# Mathematical Modeling of Routes Maintenance and Recovery Procedure for MANETs

**Zafar Iqbal†, Tahreem Saeed†, Tariq Rafiq† and Ahsan Humayun†**

†Department of Computer Science, COMSATS University Islamabad, Sahiwal Campus, Punjab, Pakistan

**Summary**
Routing is one of the most mysterious issues from the birth of networks up till now. Designing routing protocols for Mobile Ad hoc Networks (MANETs) is a complicated task because unpredictable mobility patterns of mobile nodes greatly effect routing decisions. Various routing protocols are designed to improve this very problem. Different simulator based routing protocols are designed but these protocols might fail during deployment because of the testing procedures of simulators. In this study, a novel formal model for routes management is proposed for MANETs. Formal methods are the most novel techniques based purely on mathematics and are used for the verification, validation of critical systems/models and guarantee the correctness and completeness of hardware/software systems. The proposed routing model is a complete and detailed graph based logical model defined in VDM-SL (formal language) and then verified and validated by using VDM-SL toolbox.

## 1. Introduction

Ad hoc network is a blend of wirelessly connected nodes regardless of fixed/dedicated infrastructure and is designed for some specific application according to the demand [1]. These networks do not require complex, pre-defined topology, and continuous monitoring. The most dynamic class of ad hoc networks is Mobile Ad hoc networks (MANETs) in which nodes (every communicable entity) are free to move causing a random change in the topology every time a mobile node changes its position. These networks may have mobile or static nodes connected to function remarkably in unattended and random fashion. MANETs are made flexible by implementing self-configuring capability that means whenever a node is added or removed from a network, it gets automatically registered or unregistered from the network respectively [2]. A very weak connectivity among nodes is observed in MANETs because nodes move with variable speed and follows unpredictable movement pattern causing a very random topology. Figure 1 shows the random topology of MANETs in which a rounded boundary shows the network area in which seven mobile nodes i.e. A, B, C, D, E, F, G are connected to create a network. The first circle shows the network topology at time T0 at which the nodes are connected as shown within that circle but in the second circle the network at T0 is transformed in a different topology at time T1 in which all nodes change their positions and some mobile nodes are added and some leaves the network.

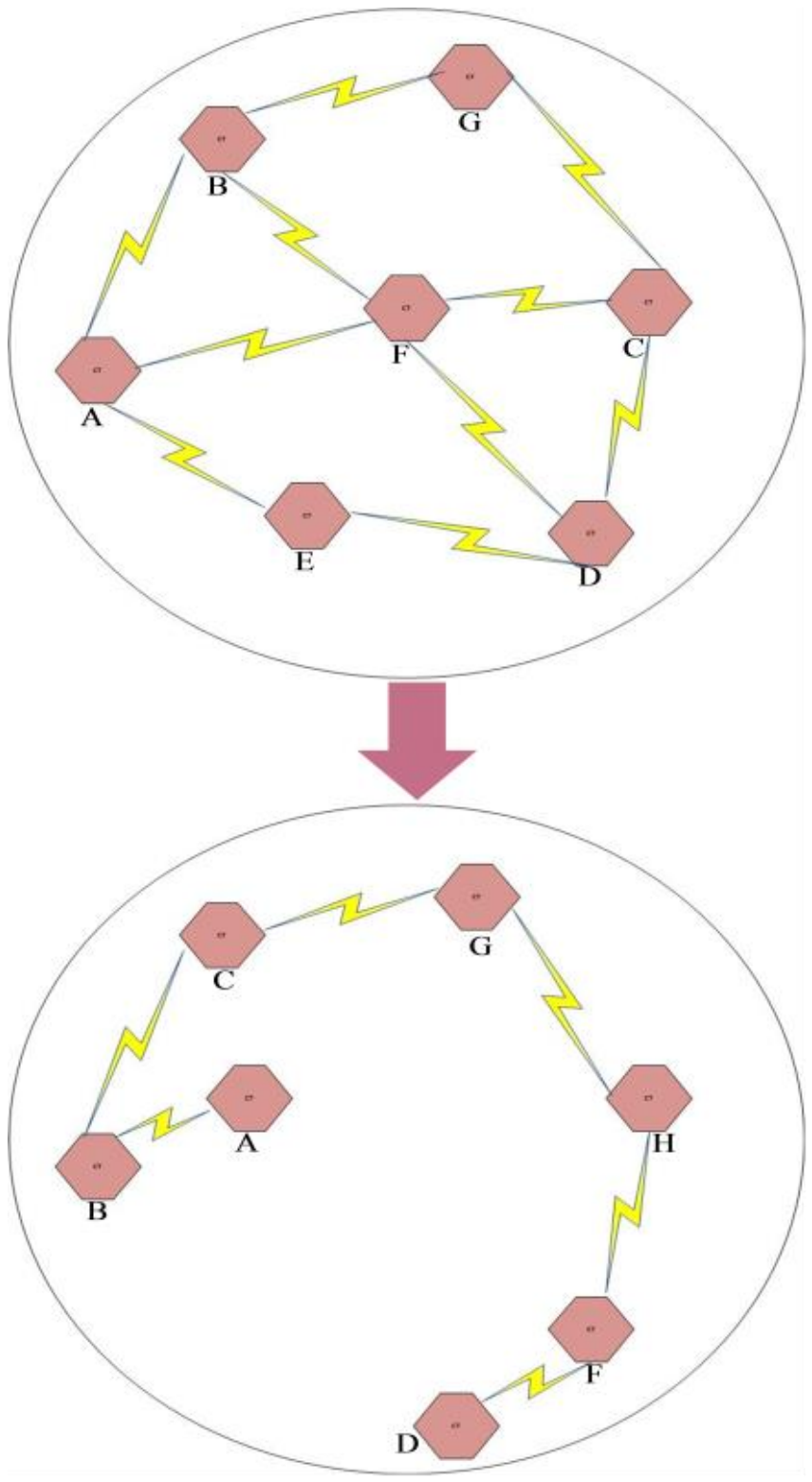

Fig. 1 MANET Topology at T0 and T1

Nodes join and leave the network according to their coverage spectrum and have very lesser and weak connections among the nodes due to their randomly changing mobility [3]. The network setup of these networks is peer to peer network architecture and no nodes have more rights than other [4]. Routing protocols for MANETs should have some special features to deal with the mobility changes, for example, it must be totally dispersed, adaptive to recurrent variations in topology, it should involve a least number of mobile nodes for route/path discovery and maintenance, and must have lowest time for connectivity set-up, should be limited to some specific area, must have no loop and self-governance against mobile routes [5]. MANETs have vast range of deployment applications where there is a need of quick and temporary network setups like military, civilian, safety critical applications etc. The criticality of these applications made MANETs topology and connectivity and more accurately said to be ‘routing’, more critical and sensitive.

There are various routing protocols designed for MANETs categorized based on the mobility model, these protocols may follow to learn route to the desired destination where there are mobile nodes. These protocols are designed to bear problems like power depletion, lower bandwidth, and high error rates, while taking routing decisions and selection of the most suitable node (termed as relay node) for routing packet from source to destination [6]. Almost all protocols for MANETs are tested and validated through simulation techniques for the realization of real-time network prediction but these techniques do not ensure the correctness and consistency of the designed protocol. There are many disasters in the past that happens due to weak testing techniques that includes vasa ship drowning that was a result of untested design and the explosion of launch vehicle Ariane 5 []. The reason behind its explosion was not following the verification results. As we have observed that MANETs are also designed for safety critical applications like military decisions, border surveillance, therefore, the topology and routing decisions of these networks are needed to be verified so that there is lesser probability of its failure. In this study, we have focused on design, testing and validation methods of routing protocols for MANETs as we have observed that this is one of the most critical issue to get addressed properly. Numerous tools are intended for the employment of graph theory based MANETs but formal methods are mathematical tools and techniques that provide the proof for verification, validation, correctness of software and hardware systems. Trend is now shifting towards verification of networks through formal methods.

In this paper, we have formally verified routing mechanism of traditional AODV protocol that is specifically designed for MANETs and introduce route recovery procedure to ensure maximum connectivity among nodes. In literature [1l9], protocol is designed by using Z notation which is a very abstract level formal language and the routings described was not sufficient for routes management in terms of bogus routes deletion and recovery. Therefore, in this paper, a complete formally verified routes management procedure is described by using VDM-SL (Vienna Development Method-Specification language) [7] because it has detailed level specification writing feature that covers almost all the functional and non-functional necessities of the proposed mode [8]. The verification of the proposed model is then done by using VDM-SL Toolbox.

## 2. Related Work

Routing protocols for MANETs are categorized in three main classes: 1) Proactive (table-driven) 2) Reactive (on-demand) and Hybrid (mixture of proactive and reactive) [9]. Proactive protocols update their routing tables unceasingly by flooding periodic routing updates to all the nodes throughout the network whereas reactive protocols update their routing tables whenever there is a demand and have no former information about the routes they may take to reach to a specific endpoint. On the other hand, hybrid protocols make responsive routing zones interrelated by proactive routing relations and regularly adjust their routing policy to the span of mobility followed by the network using reactive routing schemes [10].

A significant work has been done already for mobility management and designing routing protocols for ad hoc and sensor networks but there is a need for the settlement of this issue. Mobility-based protocols were suggested for the Ad hoc On-demand Distance Vector routing (AODV) that claim the improvement in the overall performance of the routing protocol. From route discovery to the maintenance of the routing protocols, static mobility metric was held in reserve [11]. The testing of these protocols is carried out by using simulation tools. Simulation tools generate limited number of test cases and predict the approximate results of the designed protocols but it is almost impossible to predict the dynamics of the real-time environment. Therefore, there is a need to verify the design aspect of the routing procedures by using any mathematical tools. As described in the ‘Introduction’ section, networks are now-a-days verified through formal methods before simulations. The use of formal methods increases the confidence level of the simulation results. One of the most emerging areas in this context is Mobile Ad hoc and Sensor Networks (MAHSNs). In preceding studies, it was found that a large range of uses predominantly in scheming the networks and protocols for critical areas like safety and security systems, improvement of life, environment, health, economy and energy [12]. Formal methods are broadly used in a variety of areas including modeling requirements, software engineering, simulations, and verification of network protocols, model checking, static checking, and theorem proving and for the verification of both hardware and software systems. There are more than one hundred formal languages and tools available to provide verifications, deadlock freeness and validation of a system at various development levels. Some of the languages are nearby programming languages and some deals more with the abstraction [8].

In literature [13], it is shown that a large variety of networks and routing protocols are intended by using formal methods and these are the latest and best approaches used for designing different models for MANETs. In paper [14], a di-net based routing procedure is formally verified that provides clustering/subnet based routing by using VDM-SL which is a formal specification language, described as a dynamic graph. In [15], a formal modeling of AODV based routing procedure is proposed in which a low overhead and formally verified procedure is developed for routing in MANETs. In [16], a unicast procedure is developed and formally verified for VANETs and is formalized by using VDM-SL. Literature [17,18] also shows authors interest in formalizing topological and routing aspects of networks.

Formal analysis of mobility and routings has been done using Z specification language for ad hoc networks in [19], where the mobility model and routings are described at a very abstract level because Z is a schema-based abstract formal notation. Node coverage and range based decisions are ignored and routings are not described sufficiently that is required for MANETS.

As routing decisions completely depends on the dynamic behavior of nodes in mobile ad hoc and sensor networks because it cause sudden and impulsive topological variations in the network and increases the link damage, associations and neighbor routes updates. As in literature survey [9], comparative analysis of many formal method according to specific parameters is done in which it is observed that VDM-SL is the most suitable specification language for describing networks and network protocols. In this paper, a routing protocol for MANETs is described that covers all aspects of routings from route discovery, maintenance, expiry and recovery by using VDM-SL formal language. VDM-SL ensures the detailed level description, completeness and verified correctness of the proposed model and lessens the early abstraction of the model described [19] in Z formal language.

## 3. Formal Model of Route Management

A formal detailed level model for routes management is described and proposed in this study. Routes management is the most critical issue when a network has very sudden and uncertain changes in user/mobile nodes. In this paper a framework for complete routes management is proposed by using VDM-SL (formal specification language). Past studies show that earlier route procedures are described in Z notation/language that is a very abstract language. The proposed model is a detailed level route management procedure i.e. from route initialization to maintenance and recovery. VDM-SL deals with static and dynamic properties used to describe the entire system. The VDM based proposed model comprises of types, state and necessary operations that deals with the static and dynamic structure of the model. Firstly, types are defined to describe the generic data types and then these user defined data types are used in composite data types to describe fields' data types. Text written in the shaded area below is a VDM-SL structure in which Mobile, Mid, Routestatus, Power, NodeType and ttl are defined as initial data types.

```
types
Mobile=token;
Mid = real;
Routestatus = <Activate> | <Deactivate>;
Power= <Charged> | <Low> | <Dead>;
NodeType= <Transmitter> | <Receiver> | <Intermediate>;
ttl = real;
```

After initial types definition, a composite type **mob_Range** is defined that has two fields i.e. n_range and Neighbours that is set of mobile nodes, to define the mobile nodes in coverage.

```
mob_Range :: n_range : real
             Neighbours : set of Mob_node
```

The dynamic behavior of a mobile node in MANET is described by defining a composite type. **Mob_node** is an object that has seven fields Mid, Nstaus, n_range, Seq, Neighbours and TTL to describe the node ID, node status, node range, sequence number, in range nodes and time to live (ttl) and type of a node respectively. Also a constraint of maximum 30 ms is applied on ttl, otherwise route having expired ttl will be deleted or deactivated.

```
Mob_node :: Mid : Mid
            Nstatus : Power
            n_range : mob_Range
            Seq : int
            Neighbours : set of Mobile
            TTL : ttl
            type : NodeType
inv mk_Mob_node(-,-,-,-,-,t,-) == t >= 0 and t <= 30;
```

Then, **Adhoc_con** are created and is defined as the order pair of two nodes that means that there exist a connection between that two nodes. **All_connect** is the set of these order pairs that keeps all the record of ad hoc links in it. Also an invariant is defined to ensure that no node may communicate with itself and this constraint may avoid overhead to some extent.

```
Adhoc_con = Mob_node*Mob_node;

All_connect = set of Adhoc_con
inv all_connect == forall mk_(a1,a2) in set all_connect &
mk_(a1,a2) in set all_connect and a1.Mid <> a2.Mid;
```

## 3.1 Proposed Route Modeling

The main concern and focus of this paper is to define routes and operations on routes. Basically, a route is a sequence of nodes making a complete route form source to destination. So, **Routes** are defined as sequence of mobile nodes and a composite type **rstatus** is created to define the status of routes having two fields i.e. route and status to associate a route with a status that may be activated or deactivated. For example, in Figure 2 a route from node A to node D is shown via dotted arrows in which a complete route is [A, F, B, G, C, D].

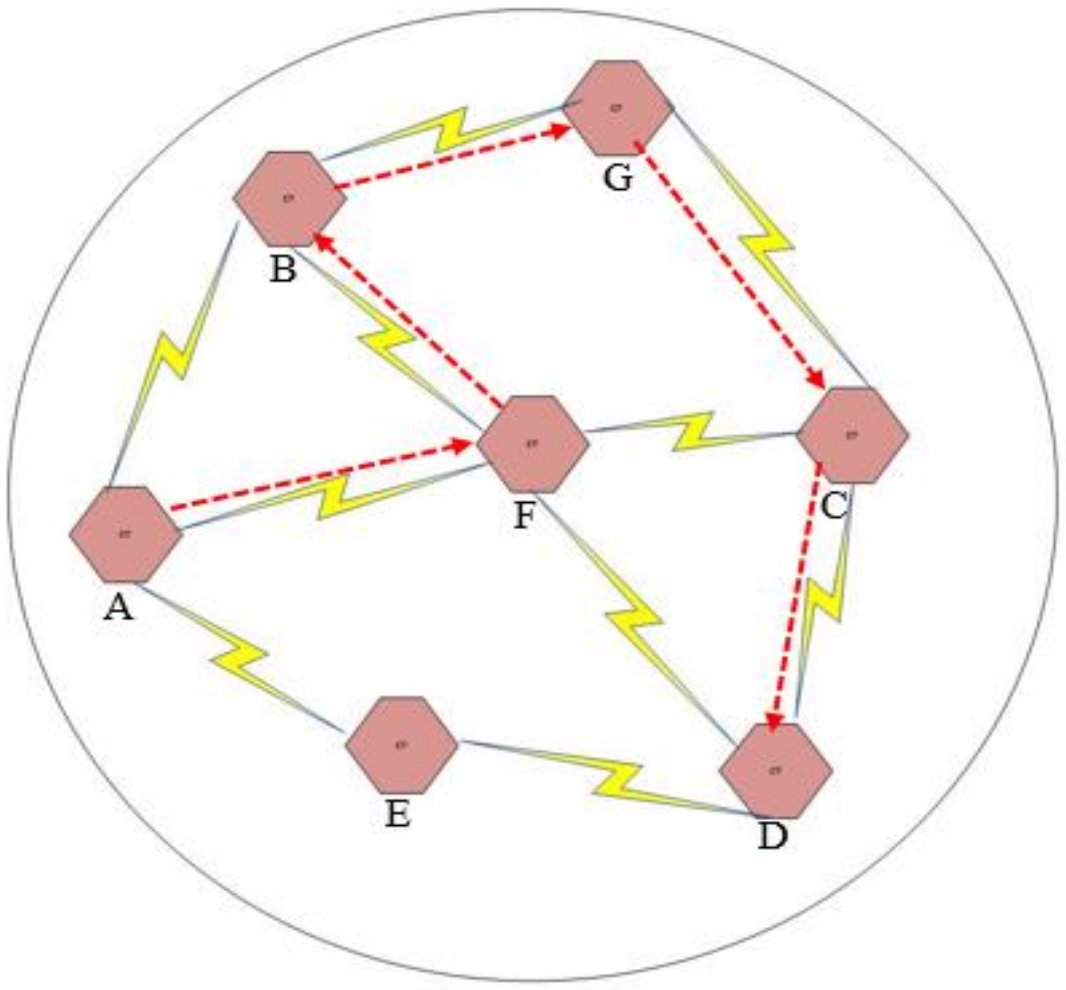

Fig. 2 Sample Route from node A to D

```
Routes = seq of Mob_node;

rstatus :: route : set of Routes
           status : Routestatus;
```

**Net_topology** is a composite type that is defined to explain the topology of the network completely in which we have Mobiles of type set of Mob_node and Connections of type All_connect. Also a constraint is defined in which the first part of the invariant says that the restriction that any two mobile nodes have linking between them and in the second part, it is proved that all the links have two nodes related with them.

```
Net_topology:: Mobiles : set of Mob_node
               Connections : All_connect
inv mk_Net_topology(Mobiles,Connections) == forall m_n1,m_n2 in set Mobiles & exists M_link in set Connections & M_link= mk_(m_n1,m_n2) and forall M_link in set Connections & exists m_n1, m_n2 in set Mobiles & M_link= mk_(m_n1,m_n2);
```

Mobile ad hoc network is a subtype of a network that is the main type of the network for which routes management model is designed/described. A type **M_adhoc** is defined that is a type of the above described network. Necessary invariants are defined that says that all nodes must have the link between them assuring the connectivity of the node, no two nodes with the same ID have links and links must be bi-directional.

```
M_adhoc= [Net_topology]
inv adhoc == forall M_link in set adhoc.Connections &
exists m_n1, m_n2 in set adhoc.Mobiles & M_link= mk_(m_n1,m_n2)
and forall m_n1, m_n2 in set adhoc.Mobiles &
exists M_link in set adhoc.Connections & M_link = mk_(m_n1,m_n2) => m_n1.Mid <> m_n2.Mid
and forall m_n1, m_n2 in set adhoc.Mobiles &
exists M_link1 in set adhoc.Connections & M_link1 = mk_(m_n1,m_n2) => (exists M_link2 in set adhoc.Connections & M_link2 = mk_(m_n2,m_n1));
```

In VDM-SL, states are used to describe the dynamic part of a system. The mobile ad hoc network must be designed in such a way that may bear all the change. **Routing** is the state of the proposed model having six fields i.e. **adhoc, routes, addedroute, deletedroute, allroutes** and **expire.**

```
state Routing of
adhoc : [M_adhoc]
routes : seq of Mob_node
addedroute : seq of Mob_node
deletedroute : seq of Mob_node
allroutes : set of Routes
expire : [rstatus]
inv mk_Routing(adhoc,routes,-,-,-,-) == forall r in set elems routes & r in set adhoc.Mobiles and len routes >= 1 => (exists m_n in set adhoc.Mobiles & (m_n = hd routes and m_n.type = <Transmitter>) ) and (exists m_n in set adhoc.Mobiles & (m_n = routes (len routes) and m_n.type = <Receiver>) ) and ((forall i in set inds routes & i >= 2 and i <= len (routes)-1) and (exists m_n in set adhoc.Mobiles & (m_n = routes (len routes) => m_n.type = <Intermediate>) ) )
and len routes > 1=> forall i in set inds routes & i >= 1 and i <= len (routes)-1 and exists m_n1,m_n2 in set adhoc.Mobiles & mk_(routes (i), routes(i+1)) = mk_(m_n1,m_n2)
init r == r = mk_Routing(nil,[],[],[],{},nil)
end
```

**State Invariant:** 1) Each node of a sequence must be a part of a network and for every non-empty sequence, the first node must be a transmitter and the last node will be a receiver and all the middle nodes of a sequence will be the intermediate nodes of a route. 2) And two consecutive nodes in a sequence must have a connection between them.

**State Initialization:** Every variable of the state is initialized according to its specified type. Here, each variable is initialized by nil, empty set and empty sequence accordingly.

Next part of the specifications comprises of the main routing operations described in detail for the specification of routing mechanism in MANETs. A keyword **'operations'** is written before defining operations. Firstly, route initialization operation is defined that describe the route initialization process initiated by the source/transmitter node.

```
operations

RouteInitiation()
ext rd adhoc : [M_adhoc]
        wr routes : seq of Mob_node
pre true
post forall m_n1 in set elems routes & m_n1.type =
<Transmitter>  and exists m_n2 in set elems routes &
m_n1.n_range = m_n2.n_range and (exists M_link1 in set
adhoc.Connections & M_link1 = mk_(m_n1,m_n2)) and
exists m_n3 in set elems routes & m_n3.type = <Receiver>
and (exists links in set adhoc.Connections & links =
mk_(m_n1,m_n3));
```

**RouteInitiation** operation i.e. defined above says that it is reading the state variables *adhoc* and then writes *routes* variable. **Pre:** No pre-constraint on any variable. **Post:** The node that initiate routes must be a transmitter node and it floods information to all the nodes that are in its range and create link between them and this procedure will be done up to the receiver node.

```
RouteAddition()
ext rd adhoc : [M_adhoc]
        wr allroutes : set of Routes
        rd addedroute : seq of Mob_node
pre true
post forall m_n1 in set elems addedroute & exists m_n2 in
set adhoc.Mobiles & m_n2 = m_n1 and exists m_n1,m_n2
in
set adhoc.Mobiles & len addedroute >= 1 => hd addedroute
= m_n1 and addedroute(len addedroute) = m_n2 and
m_n1.type = <Transmitter> and m_n2.type = <Receiver>
and allroutes = allroutes union {addedroute};
```

**RouteAddition** operation defined to operate in a case if we have a new route created and it must be inserted in all routes. So, this operation has three variables i.e. *adhoc, allroutes* and *addedroutes* to be read or written. **Pre:** No pre-constraint on variables. **Post:** All the nodes of the new sequence/route must be a part of the network and the first and last node of the new route must be a transmitter and receiver respectively and new route will be added to the set of all routes by taking the union of the old set and the new sequence.

```
ReverseRoute()
ext rd adhoc : [M_adhoc]
        wr routes : seq of Mob_node
pre true
post forall m_n1 in set elems routes & m_n1.type =
<Receiver>  and exists m_n2 in set elems routes &
m_n1.n_range = m_n2.n_range and (exists M_link1 in set
adhoc.Connections & M_link1 = mk_(m_n1,m_n2)) and
exists m_n3 in set elems routes & m_n3.type =
<Transmitter> and (exists links in set adhoc.Connections &
links = mk_(m_n1,m_n3));
```

**ReverseRoute** operation is defined to describe the reply back procedure from the destination to the source node. The above-mentioned operation has two variables like route initiation operation. **Pre:** No restriction on any of the variable. **Post:** The receiver node will create connection to all its predecessor neighbors and instead of flooding; it will create a uncast/unique link from receiver to the transmitter.

```
RouteExpiry()
ext rd adhoc : [M_adhoc]
        rd allroutes : set of Routes
        wr expire : [rstatus]
pre true
post forall r in set expire.route & exists m_n in set elems r &
m_n.TTL >= 30 => expire.status = <Deactivate>;
```

**RouteExpiry** procedure is defined to explain the expired routes' logic that is if the route has any node with the expired time to live then that link/route will be deactivated. **Pre:** No restriction. **Post:** If any node has expired TTL then its status will be set to deactivate.

```
RouteDeletion()
ext rd adhoc : [M_adhoc]
        wr allroutes : set of Routes
        rd deletedroute : seq of Mob_node
pre true
post forall m_n1 in set elems deletedroute & exists m_n2 in
set adhoc.Mobiles & m_n2 = m_n1 and exists m_n1,m_n2
in set adhoc.Mobiles & len deletedroute >= 1 => hd
deletedroute = m_n1 and deletedroute(len deletedroute) =
m_n2
and m_n1.type = <Transmitter> and m_n2.type =
<Receiver> and allroutes = allroutes \ {deletedroute};
```

**RouteDeletion** operation is defined to operate in a case if we have any unnecessary route to be deleted then this operation will be applied. **Pre:** No pre- constraint on variables. **Post:** All the nodes of the sequence/route to be deleted must be a part of the network and the first and last node of the new route must be a transmitter and receiver respectively and new route will be deleted from the set of all routes by taking the difference of the old set and the new sequence.

```
RouteRecovery()
ext rd adhoc : [M_adhoc]
        rd allroutes : set of Routes
        wr expire : [rstatus]
pre forall r in set expire.route & exists m_n in set elems r &
m_n.Nstatus = <Dead> or expire.status = <Deactivate>
post expire.status = <Activate>;
```

**RouteRecovery** procedure as defined above is applicable when any deactivated route will be activated. Pre: If there exist any expired routes that have any node with dead status and any route with the deactivated status in set of all routes. Post: The status of that route will be set to be activated.

## 4. Result and Analysis

In this paper, a formal analysis and modeling of the model proposed for routes management for MANETs is done through VDM-SL toolbox. As described previously that formal modeling and analysis is required for the detailed examination of any critical system/model, therefore, during model analysis in VDM-SL, analysis shows that there were some data type and integrity property violations. Hence, constraints were applied on state and operations in shape of invariants and pre/post conditions to remove all the bugs in the proposed model. Then after application of all these properties, no syntax, type errors and no warnings were found after the complete analysis of the proposed routing model. Few errors were confronted during write up phase of specifications but those were fixed at modeling time by refining the constraints and properties on the variables. Invariants on different variables define the precision of the model. Figure 3 depicts the screenshots that provides the proof of the correctness of the formal modeling of the routing protocol in which S and T shows that there are no syntax and type errors in the mathematically specified routing model and C shows that C++ code file of the complete routing protocol is automatically generated by the tool that might be used for further designing of any network by using any (IDE) Integrated Development Environment. Table 1 shows all property checks against all properties of the designed model.

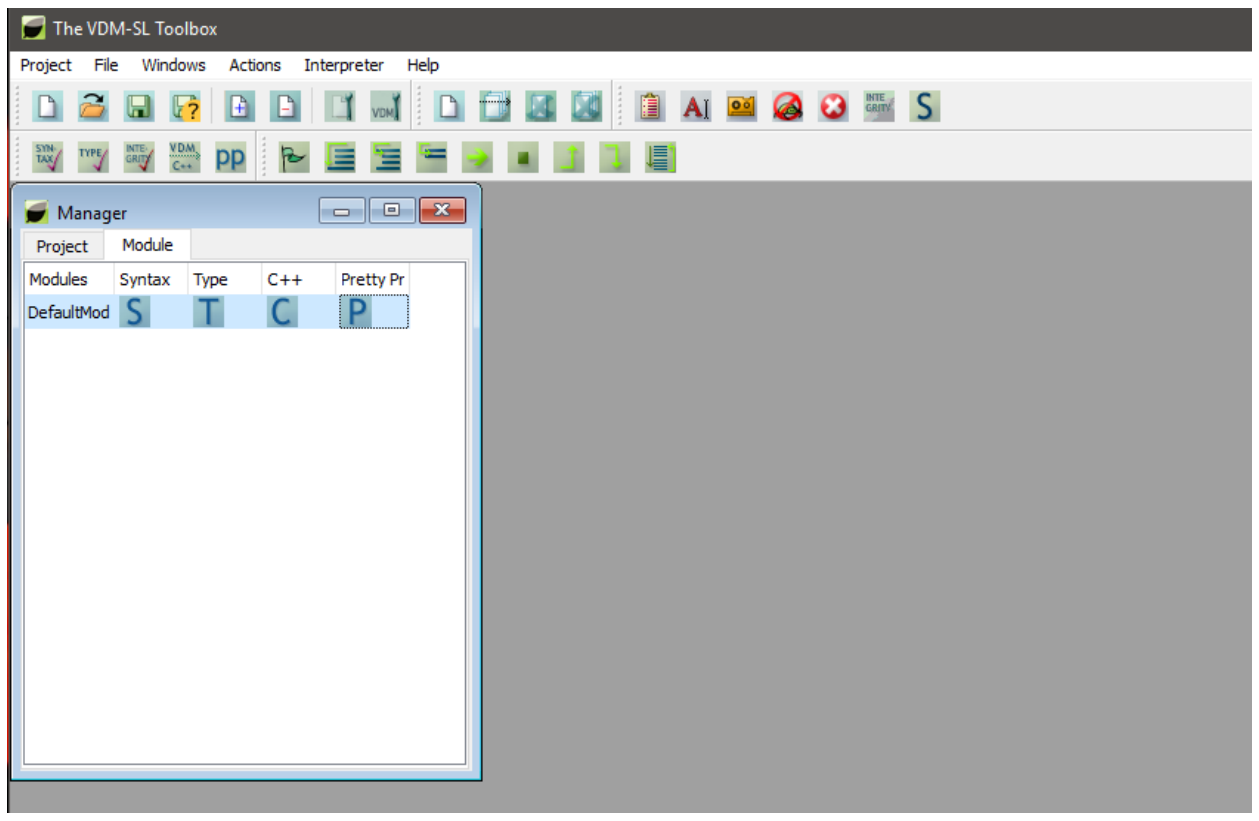

Fig. 3 Proof of Correctness

Table 1: VDM-SL Analysis Report

| Property Name | Syntax Checker | Type Checker | Code Generator | Pretty Printer |
|---|---|---|---|---|
| Primitive Data types e.g. nodes, links etc. | ✓ | ✓ | ✓ | ✓ |
| Network Topology | ✓ | ✓ | ✓ | ✓ |
| State | ✓ | ✓ | ✓ | ✓ |
| Operation e.g. Routes Initialization | ✓ | ✓ | ✓ | ✓ |

## 5. Conclusion

In this paper, we have proposed a formally verified routing model for mobile ad hoc networks. Existing routing protocols/schemes are purely simulation based models and does not promise the completeness and correctness of the models because simulator based testing have fewer, selective and random test cases and are mostly assumption based inputs. In this study, innovative and embryonic methodology i.e. formal methods are used to specify the complete specification of the proposed routing model. In literature, Z method specification of the routing protocol is described at a very abstract level and routes initialization, expiry and recovery procedures are defined. Therefore, there is a need for detailed level formally verified and specified complete routes management protocol that specifies all the deficiencies of existing models. The proposed model is a VDM-SL based detailed level specified mobility model that is more efficiently designed and all the routing decisions will be according to the specified procedures defined for route discovery up to its recovery. The model is then analyzed through VDM-SL toolbox that then confirm the correctness, consistency, verification and validation of the proposed routing scheme/model for MANETs.

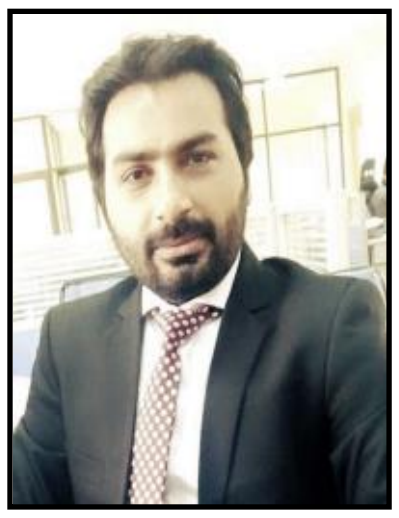

**Zafar Iqbal** is pursuing his PhD at COMSATS University Islamabad, Pakistan. Currently, he is serving as Assistant Professor at COMSATS University Islamabad, Sahiwal Campus, Punjab, Pakistan. He received his MS Computer Science in 2008 from University of Engineering and Technology (UET), Lahore, Pakistan. He qualified CISCO Certified Academic Instructor (CCAI) exam with distinction. He also worked in High Performance Computing Lab (HPCL) at Al-Khwarizmi Institute of Computer Science (KICS), UET, Lahore during his master studies. He has supervised various ICT R&D funded projects. Now, he has great enthusiasm towards his research activities. His current research interests include Wireless Ad hoc Networks and its applications, Internet of Things, Internet of Vehicles, and Modeling of Critical Systems using Formal Approaches.

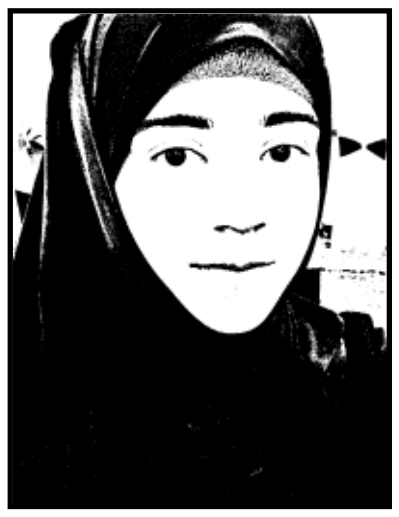

**Tahreem Saeed** received MS in Computer Science degree in 2017, from COMSATS Institute of Information Technology (CIIT), Sahiwal, Pakistan. She is currently serving as a Lecturer in Department of Computer Science at COMSATS University Islamabad, Sahiwal Campus, Punjab, Pakistan. She received her BS degree in Telecommunication and Networking with distinction of Gold Medal from the same prestigious Institute. She won ICT R&D funding for her research project in BSTN. Her research interests are Wireless Ad hoc networks and its applications, Internet of things, IoV, IoMT, Big Data, Modeling of Systems using Formal Approaches, Various applications and Modeling of Biosensor Antennas.

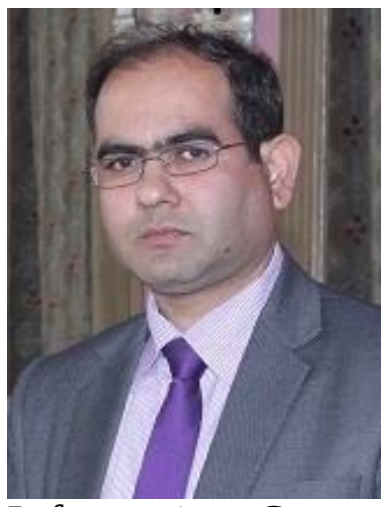

**Tariq Rafiq** is currently working as Lecturer at Department of Computer Science, COMSATS University Islamabad, Sahiwal Campus, Pakistan. He received MS degree in Computer Science from Virtual University of Pakistan in 2016. Now he is pursuing his PhD from COMSATS University Islamabad. His research interests include Formal Modeling, Information Systems, Database, Data Warehouse, and Natural Language Processing

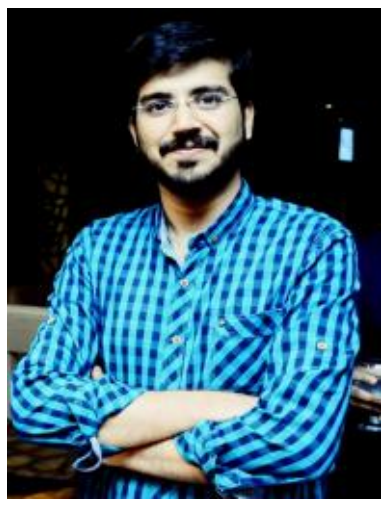

**Ahsan Humayun** received the B.S (2014) and M.S (2017) degrees in Computer Science from National Textile University Faisalabad, Pakistan. He joined University of Central Punjab, Faisalabad Campus in 2015 as a Lecturer in Department of Computer Science. Since April 2018, He has been working as a Lecturer in Dept. of Computer Science at COMSATS University Sahiwal Campus, Pakistan. He has published 3 International Journal publications. His main research areas are High Performance Computing, Usage of Data mining techniques and Wireless sensor networks in Healthcare sector.